\documentclass[epj,twocolumn]{webofc}
\usepackage[varg]{txfonts}   
\woctitle{Flavour changing and conserving processes}

\usepackage{hyperref}
\usepackage{mathtools,siunitx,booktabs,braket,slashed}
\usepackage{tikz}
\usepackage[export]{adjustbox} 

\usepackage[percent]{overpic}

\newcommand{\be}{\begin{equation}}
\newcommand{\ee}{\end{equation}}
\newcommand{\bea}{\begin{eqnarray}}
\newcommand{\eea}{\end{eqnarray}}

\DeclarePairedDelimiter\abs{\lvert}{\rvert}

\newcommand\amuhlbl{a_\mu^{\text{HLbL}}}
\newcommand\amulbl{a_\mu^{\text{LbL}}}
 
\renewcommand{\>}{\rangle}  
\newcommand\wtg[1]{\bar{\mathfrak{g}}^{(#1)}}

\newcommand{\fm}{\mathrm{fm}}
\newcommand{\MeV}{\mathrm{MeV}}
\newcommand{\GeV}{\mathrm{GeV}}

\newcommand{\FF}{{\cal F}_{\pi^0\gamma^*\gamma^*}}

\newcommand{\LMDV}{\mathrm{LMD+V}}
\newcommand{\dof}{\mathrm{d.o.f.}}

\newcommand{\amupi}{a_\mu^{\mathrm{HLbL}; \pi^0}}

\newcommand{\GammaGG}{\Gamma_{\gamma\gamma}}

\newcommand{\MsTT}{\overline{\mathcal{M}}_{TT}}
\newcommand{\MsTTa}{\overline{\mathcal{M}}_{TT}^{a}}
\newcommand{\MsTTt}{\overline{\mathcal{M}}_{TT}^{\tau}}
\newcommand{\MsTL}{\overline{\mathcal{M}}_{TL}}
\newcommand{\MsLT}{\overline{\mathcal{M}}_{LT}}
\newcommand{\MsTLt}{\overline{\mathcal{M}}_{TL}^{\tau}}
\newcommand{\MsTLa}{\overline{\mathcal{M}}_{TL}^{a}}
\newcommand{\MsLL}{\overline{\mathcal{M}}_{LL}}

\makeatletter
\g@addto@macro\bfseries{\boldmath}
\g@addto@macro\normalfont{\unboldmath}
\makeatother

\begin{document}

\selectlanguage{english}

\title{Hadronic light-by-light scattering contribution to the muon
  $g-2$ on the lattice}


\author{%
\firstname{Nils} \lastname{Asmussen}\inst{1} \and 
\firstname{Antoine} \lastname{G\'erardin}\inst{2} \and 
\firstname{Jeremy} \lastname{Green}\inst{3} \and
\firstname{Oleksii}  \lastname{Gryniuk}\inst{1} \and
\firstname{Georg}  \lastname{von Hippel}\inst{1} \and
\firstname{Harvey B.}  \lastname{Meyer}\inst{1,2} \and \\ 
\firstname{Andreas}~\lastname{Nyffeler}\inst{1}\fnsep\thanks{Speaker,
  \email{nyffeler@uni-mainz.de} } \and
\firstname{Vladimir}  \lastname{Pascalutsa}\inst{1} \and
\firstname{Hartmut}  \lastname{Wittig}\inst{1,2} 
}


\institute{%
PRISMA Cluster of Excellence and Institut f\"ur Kernphysik, 
Johannes Gutenberg-Universit\"at Mainz, D-55099 Mainz, Germany 
\and
Helmholtz Institut Mainz, D-55099 Mainz, Germany
\and
John von Neumann Institute for Computing (NIC), DESY, Platanenallee 6,
D-15738 Zeuthen, Germany 
}


\abstract{%
  We briefly review several activities at Mainz related to hadronic
  light-by-light scattering (HLbL) using lattice QCD. First we present
  a position-space approach to the HLbL contribution in the muon
  $g-2$, where we focus on exploratory studies of the pion-pole
  contribution in a simple model and the lepton loop in QED in the
  continuum and in infinite volume. The second part describes a
  lattice calculation of the double-virtual pion transition form
  factor $\FF(q_1^2, q_2^2)$ in the spacelike region with photon
  virtualities up to $1.5~\GeV^2$ which paves the way for a lattice
  calculation of the pion-pole contribution to HLbL.  The third topic
  involves HLbL forward scattering amplitudes calculated in lattice
  QCD which can be described, using dispersion relations (HLbL sum
  rules), by $\gamma^*\gamma^* \to \mbox{hadrons}$ fusion cross
  sections and then compared with phenomenological models.  }

\maketitle


\section{Introduction}
\label{intro}

The anomalous magnetic moment of the muon has served for many years as
a precision test of the Standard Model~\cite{JN_09, Jegerlehner_15_17,
  Knecht, Laporta} and it has also played an important role in many
presentations at this meeting. There is a discrepancy of $3-4$
standard deviations between experiment and theory for some time
now. This could be a signal of New Physics~\cite{JN_09, Stoeckinger},
but the uncertainties in the theory prediction from hadronic vacuum
polarization (HVP) and hadronic light-by-light scattering (HLbL) make
it difficult to draw firm conclusions. In view of upcoming four-fold
more precise new experiments at Fermilab and
J-PARC~\cite{Fermilab_J-PARC}, these hadronic contributions need to be
better controlled.

The improvement for HVP looks straightforward, with more precise
experimental data from various experiments on hadronic cross-sections
as input for a dispersion relation~\cite{HVP_DR}. But also lattice QCD
is getting more and more precise~\cite{HVP_Lattice}, and, hopefully,
also a new method using muon-electron scattering to measure the
running of $\alpha$ and the HVP in the spacelike
region~\cite{HVP_spacelike}, will be feasible at some point with the
required precision. 

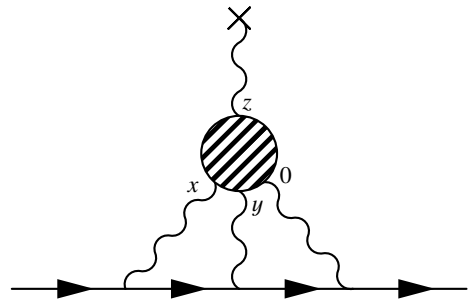
\begin{figure}[b!] 
   \centerline{%
      \begin{overpic}[width=0.36\textwidth]{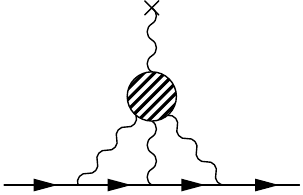}
         \put(39,24){\small $x$}
         \put(53,20){\small $y$}
         \put(59,26){\small $0$}
         \put(51,42){\small $z$}
      \end{overpic}
   }
   \caption{HLbL scattering contribution to the muon $g-2$.}
   \label{fig:muonhlbl}
\end{figure}

On the other hand, the HLbL contribution to the muon $g-2$, see
Fig.~\ref{fig:muonhlbl}, has only been calculated using models so
far~\cite{JN_09, Knecht, Bijnens_16_17} and the frequently used
estimates from Refs.~\cite{PdeRV_09, N_09, JN_09} (revised slightly in
Ref.~\cite{Jegerlehner_15_17}) suffer from uncontrollable
uncertainties. In view of this, dispersion relations have been
proposed a few years ago~\cite{HLbL_DR_Bern_Bonn, HLbL_DR_Mainz,
  Danilkin_Vanderhaeghen_17} (see also the very recent new proposal in
Ref.~\cite{Hagelstein_Pascalutsa}) to determine the presumably
numerically dominant contributions from a single neutral pion-pole
(light pseudoscalar-pole) and from the two-pion intermediate state
(pion-loop) based on input from experimental data for the dispersion
relations. Still some modelling will be needed to estimate the
contributions from multi-pion intermediate states, like the
axial-vector contribution.

Finally, lattice QCD was proposed some time ago as a
model-independent, first principle approach to the HLbL contribution
in the muon $g-2$ and some promising progress has been achieved
recently by the RBC-UKQCD collaboration~\cite{HLbL_Lattice_RBC-UKQCD}.

Independently, also the lattice group at Mainz has studied HLbL in
recent years. We used complementary approaches to tackle the full HLbL
contribution in the muon $g-2$ with a new position-space
approach~\cite{HLbL_Lattice_Mainz_15, HLbL_Lattice_Mainz_16,
  HLbL_Lattice_Mainz_17}, studied the pion transition form factor
(TFF) with two virtual photons on the lattice to evaluate the
pion-pole contribution to HLbL~\cite{TFF_Lattice} and we analyzed HLbL
forward scattering amplitudes that can be compared using dispersion
relations (HLbL sum rules)~\cite{HLbL_sum_rules} to phenomenological
models for photon fusion processes~\cite{HLbL_SR_Lattice_15,
  HLbL_Lattice_Mainz_15, HLbL_SR_Lattice_17}. These three topics will
be discussed in the following three sections. For all the details, we
refer to the quoted papers.

\section{Position-space approach to HLbL in the muon $g-2$ on the lattice} 

The HLbL scattering contribution to the anomalous magnetic moment of
the muon in Fig.~\ref{fig:muonhlbl} can be split into a perturbative
QED kernel $\bar{\mathcal L}$ that describes the muon and photon
propagators and a non-perturbative QCD four-point function
$i\widehat\Pi$ (denoted by a blob in the Feynman diagram) that will be
evaluated on the lattice. 

The projection on the muon $g-2$ yields the master formula (in
Euclidean space
notation)~\cite{HLbL_Lattice_Mainz_15,HLbL_Lattice_Mainz_16}
\be \label{master}  
   \amuhlbl = \!\!\frac{m e^6}{3} \!\!\int \!\!d^4y \!\!\int \!\!d^4x 
      \underbrace{\bar{\cal L}_{[\rho,\sigma];\mu\nu\lambda}(x,y)}_{\rm QED}   
      \underbrace{i\widehat\Pi_{\rho;\mu\nu\lambda\sigma}(x,y)}_{\rm QCD},   
\ee 
with the spatial moment of the four-point function
\be 
   i\widehat \Pi_{\rho;\mu\nu\lambda\sigma}(x,y)  = - \int
   \!\!d^4z \, z_\rho \, \<j_\mu(x) j_\nu(y) j_\sigma(z) j_\lambda(0)\>.
\ee 

We evaluate the QED kernel in the continuum and in infinite volume and
thereby avoid $1/L^2$ finite-volume effects from the massless
photons. Since Lorentz covariance is manifest in our approach, the
eight-dimensional integral in Eq.~(\ref{master}) can be reduced, after
contracting all indices, to a three-dimensional integral over the
Lorentz invariants $x^2, y^2$ and $x \cdot y$.

\begin{figure*}[t!]
\centering 
\resizebox{\textwidth}{!}{
\begin{tabular}{rr}
         \includegraphics[valign=b]{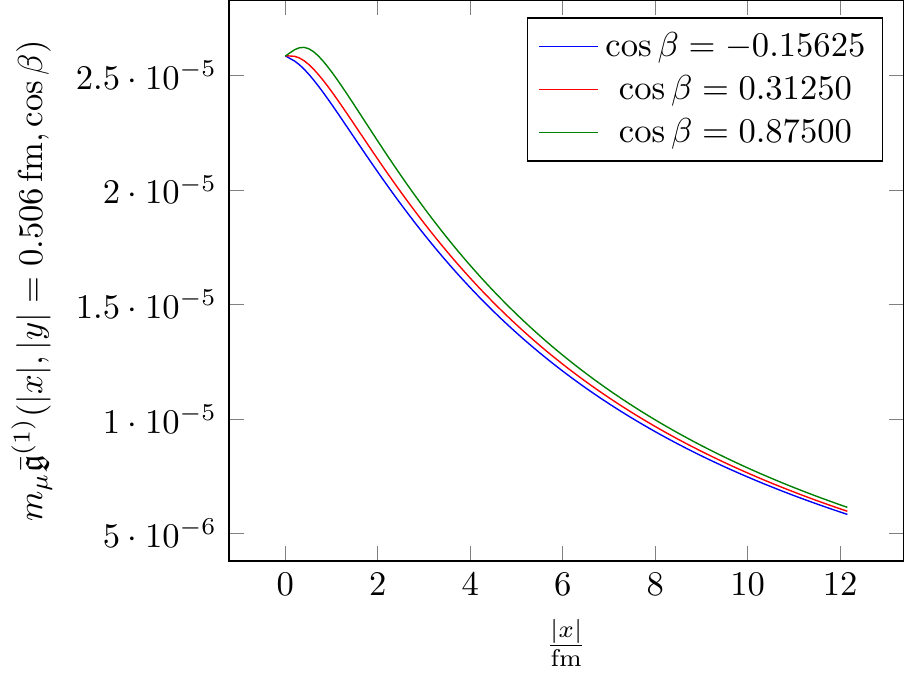} & 
         \includegraphics[valign=b]{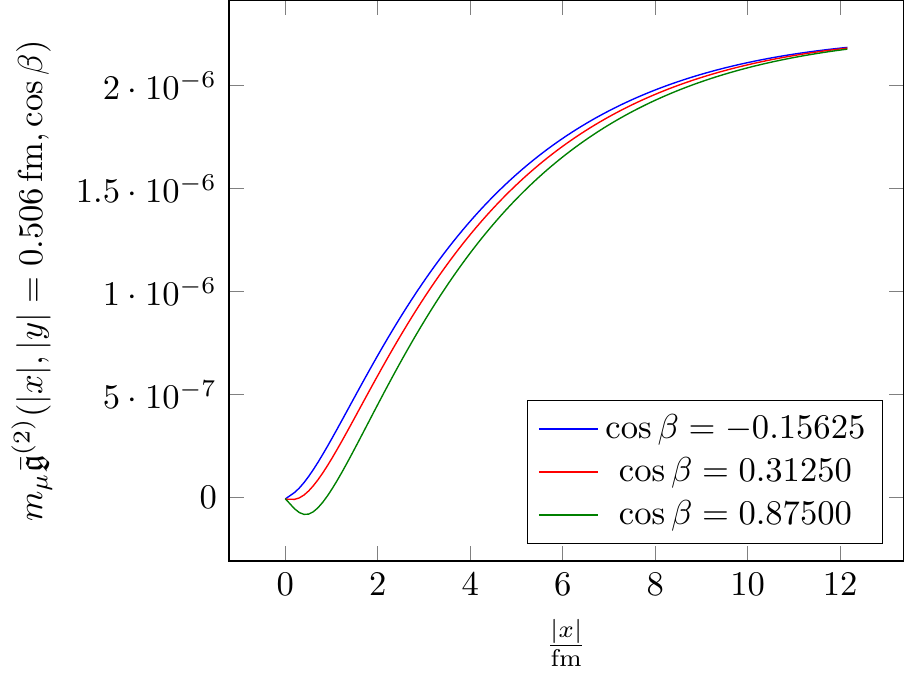}
\end{tabular} 
}
         \caption{The $\abs x$ dependence of the weight functions
           $\bar{\mathfrak{g}}^{(1)}$ (left) and
           $\bar{\mathfrak{g}}^{(2)}$ (right) for $\abs
           y=\SI{0.506}{fm}$ and three values of $\cos\beta$. Apart
           from the scale, the weight functions
           $\bar{\mathfrak{g}}^{(0)}, \bar{\mathfrak{l}}^{(1)}$ and
           $\bar{\mathfrak{l}}^{(2)}$ have the same shape as
           $\bar{\mathfrak{g}}^{(1)}$, whereas
           $\bar{\mathfrak{l}}^{(2)}$ looks similar to
           $\bar{\mathfrak{g}}^{(2)}$.}
   \label{fig:weights}
\end{figure*}

The QED kernel $\bar{\cal L}_{[\rho,\sigma];\mu\nu\lambda}(x,y)$ can
be decomposed into several tensors 
\be
\bar {\cal L}_{[\rho,\sigma];\mu\nu\lambda}(x,y) 
= \sum_{A={\rm I,II,III}} {\cal
  G}^A_{\delta[\rho\sigma]\mu\alpha\nu\beta\lambda}
T^{(A)}_{\alpha\beta\delta}(x,y).  
\ee
The  ${\cal G}^A_{\delta[\rho\sigma]\mu\alpha\nu\beta\lambda}$ are
traces of gamma matrices and just yield sums of products of Kronecker
deltas. The tensors $T^{(A)}_{\alpha\beta\delta}$ are decomposed into
a scalar~$S$, vector~$V$ and tensor~$T$ part 
\bea
\label{eq:TI}
T^{({\rm I})}_{\alpha\beta\delta}(x,y) &=&   \partial^{(x)}_\alpha
(\partial^{(x)}_\beta + \partial^{(y)}_\beta) V_\delta(x,y),
\\
\label{eq:TII}
T^{({\rm II})}_{\alpha\beta\delta}(x,y) &=& 
m \partial^{(x)}_\alpha 
\Bigg( T_{\beta\delta}(x,y) + \frac{1}{4}\delta_{\beta\delta} S(x,y)\Bigg),
\\
\label{eq:TIII}
T^{({\rm III})}_{\alpha\beta\delta}(x,y) &=&  m (\partial^{(x)}_\beta
+ \partial^{(y)}_\beta) \Big( T_{\alpha\delta}(x,y) \Big. \nonumber \\
& & \qquad \qquad \quad \Big. 
+ \frac{1}{4}\delta_{\alpha\delta} S(x,y)\Big).  
\eea
These can be parametrized by six weight functions
\bea
   S(x,y) &=& \wtg{0}, \phantom{\frac{1}{1}}
   \\
   V_\delta(x,y)
   &=& x_\delta \bar{\mathfrak{g}}^{(1)} 
     + y_\delta \bar{\mathfrak{g}}^{(2)}, 
   \\
    T_{\alpha\beta}(x,y) &=& (x_\alpha x_\beta - 
      \frac{x^2}{4}\delta_{\alpha\beta})\; \bar{\mathfrak{l}}^{(1)}
      \nonumber \\ 
   & & + (y_\alpha y_\beta - \frac{y^2}{4}\delta_{\alpha\beta})\;
   \bar{\mathfrak{l}}^{(2)} \nonumber \\ 
   & & + (x_\alpha y_\beta + y_\alpha x_\beta - \frac{x\cdot
     y}{2}\delta_{\alpha\beta})\; \bar{\mathfrak{l}}^{(3)}, 
\eea 
that depend on the three variables $x^2$, $x \cdot y = |x| |y|
\cos\beta$ and $y^2$. The semi-analytical expressions for the weight
functions have been precomputed to about 5 digits precision and stored
on a three-dimensional grid. For illustration, the expression for the
weight functions $\bar{\mathfrak{g}}^{(2)}$ has been given in
Ref.~\cite{HLbL_Lattice_Mainz_16}. In Fig.~\ref{fig:weights} we show
the two weight functions $\bar{\mathfrak{g}}^{(1)}$ and
$\bar{\mathfrak{g}}^{(2)}$ as a function of $|x| < 12~\fm$, for a
fixed value of $|y| = 0.506~\fm$ and three values of
$\cos\beta$. Plots for all six weight functions can be found in
Ref.~\cite{HLbL_Lattice_Mainz_17}.

To test our semi-analytical expressions and the software for the QED
kernel, we have computed the $\pi^0$-pole contribution to HLbL in a
simple vector-meson domi\-nance (VMD) model as well as the lepton-loop
contribution to $\amulbl$ in QED, where the results are well known. To
this aim, we first derived analytical expressions for these
contributions to the four-point function $i\widehat
\Pi_{\rho;\mu\nu\lambda\sigma}(x,y)$ in position-space, see
Ref.~\cite{HLbL_Lattice_Mainz_17} for details.

\begin{figure*}[t!]
\centering 
\resizebox{\textwidth}{!}{
\begin{tabular}{rr}
   \includegraphics[valign=b]{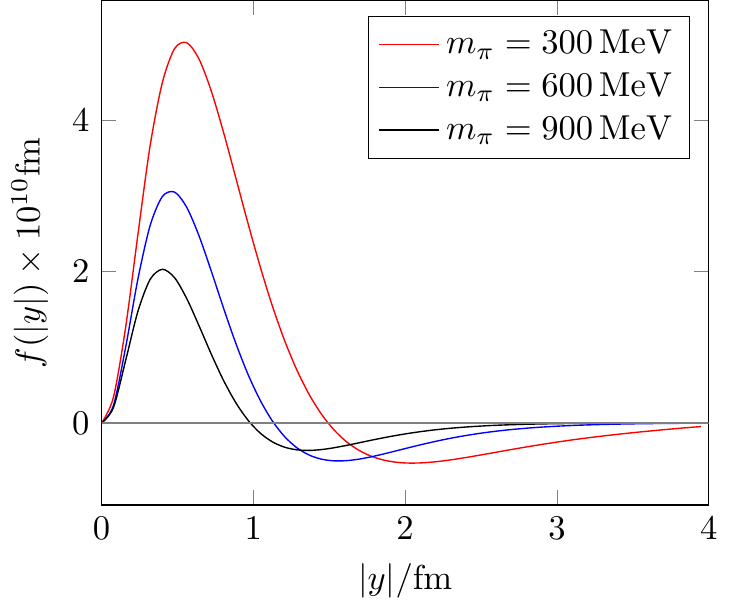} & 
   \includegraphics[valign=b]{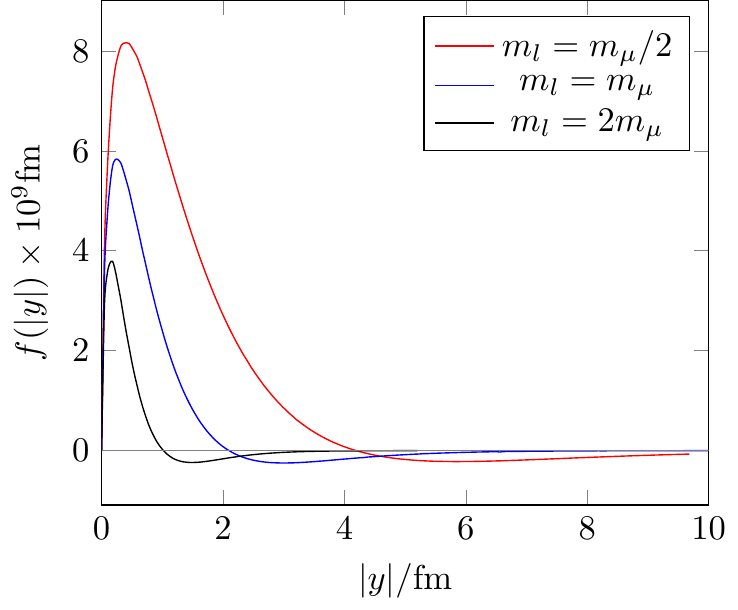}
\end{tabular} 
}
      \caption{(Left) Integrand of the pion-pole contribution
        $\amuhlbl$ for a simple VMD transition form factor for three
        different values of the pion mass. (Right) Integrand of the
        lepton-loop contribution $\amulbl$ in QED for three different
        lepton masses $m_l$ in the loop.} 
   \label{fig:integrands}
\end{figure*}

In figure~\ref{fig:integrands} we plot the integrand $f(|y|)$ of the
final integration over $|y|$ of the HLbL contribution $\amuhlbl =
\int_0^\infty d|y| \, f(|y|)$ (LbL in QED) for these two examples, after
contracting all Lorentz indices and the integrations over $|x|$ and
$\cos\beta$ have been performed in Eq.~(\ref{master}).

From the plot of the integrand for the pion-pole contribution in
Fig.~\ref{fig:integrands} one observes that this contribution to
$\amuhlbl$ is remarkably long-range with a long negative tail at large
$|y|$. One expects an exponential decay $\sim e^{- \tilde c m_\pi
  |y|}$ of the correlation function. But this seems to be countered by
some non-negligible power-like behavior $|y|^n$. For pion masses
$m_\pi = 300-900~\MeV$ we reproduce the known results, which can be
easily obtained from the three-dimensional integral representation in
momentum space given in Ref.~\cite{JN_09}, at the percent level. On
the other hand, for the physical pion mass, one will need rather large
lattices of the order of $5-10~\fm$ to capture the negative tail at
large $|y|$ in a QCD lattice simulation. Hopefully we can correct for
finite-size effects on this contribution, by computing the relevant
neutral pion transition form factor on the same lattice ensembles, see
Ref.~\cite{TFF_Lattice} and Section~\ref{sec:Pion_TFF}.

For the lepton-loop in QED the behavior of the integrand for small
$|y|$ is compatible with $f(|y|) \propto |y| \log^2(|y|)$.  This is
quite steep and means that we probe the QED kernel at small
distances. In addition, the height of the positive peak grows with
smaller masses $m_l$ of the lepton in the loop.  Furthermore there is
again a long negative tail at large $|y|$ which demands the use of a
large size for the grid where the weight functions have been
calculated, in particular for a lepton in the loop that is lighter
than the muon. For $m_l = m_\mu, 2 m_\mu$ we reproduce the
analytically known results for $\amulbl$ in QED~\cite{Lepton_loop} at
the percent level, see Table~\ref{tab:leptonloop}. On the other hand,
for the lightest lepton mass $m_l = m_\mu / 2$, some further
refinements of our numerical evaluation are needed. Once this is
achieved, we plan to make the QED kernel publicly available.

\begin{table}[b!] 
\centering 
\caption{Results $(\times 10^{11})$, precision and deviation for the
  lepton-loop contribution to LbL in QED with our approach compared to
  the known results~\cite{Lepton_loop}. The first uncertainty
  originates from the three-dimensional numerical integration, the
  second from the extrapolation of the integrand to small $|y|$.}    
   \label{tab:leptonloop}
{\small 
      \begin{tabular}{cSr@{}lll}
         \hline 
           $m_l / m_\mu$
         & {$a_\mu^{\text{LbL}}$ \cite{Lepton_loop}}
         & \multicolumn{2}{c}{$a_\mu^{\text{LbL}}$}
         & {Prec.}
         & {Dev.}
         \\ 
         \hline 
         1/2 & 1229.07 & 1257.5&(6.2)(2.4) & 0.5\% & 2.3\%  \\  
         1   &  464.97 &  470.6&(2.3)(2.1) & 0.7\% & 1.2\%  \\
         2   &  150.31 &  150.4&(0.7)(1.7) & 1.2\% & 0.06\% \\
         \hline 
      \end{tabular}
}
\end{table}

\section{Lattice calculation of the pion transition form factor
  $\FF(q_1^2, q_2^2)$}
\label{sec:Pion_TFF}

The pion transition form factor can be defined from the following
correlation function in Euclidean space, using the methods proposed
and used before in Ref.~\cite{Method_TFF}
\bea 
M^{E}_{\mu\nu}(p, q_1) & = & - \int \mathrm{d} \tau \, e^{\omega_1
  \tau} \int \mathrm{d}^3 z \, e^{-i \vec{q}_1 \vec{z}} \, \nonumber \\
&& \quad \times \, \langle 0 | T \left\{ J_{\mu}(\vec{z}, \tau)
  J_{\nu}(\vec{0}, 0) \right\} | \pi(p) \rangle  \nonumber \\  
& = & \epsilon_{\mu\nu\alpha\beta} \, {q_1}_{\alpha} \, {q_2}_{\beta} \ 
\mathcal{F}_{\pi^0\gamma^*\gamma^*}(q_1^2, q_2^2), 
\label{definition_TFF}
\eea 
provided the photon virtualities satisfy $q_{1,2}^2 < {\rm
  min}(M_{\rho}^2, 4 m_{\pi}^2)$ to avoid poles in the analytical
continuation from the original definition of the form factor in
Minkowski space to Euclidean space. The free real parameter $\omega_1$
denotes the zeroth (energy) component of the four-momentum $q_1 =
(\omega_1, \vec{q}_1)$.

The main object to compute on the lattice is the three-point function    
\bea 
\lefteqn{C^{(3)}_{\mu\nu}(\tau,t_{\pi}; \vec{p}, \vec{q}_1,\vec{q}_2)}
\label{lat_cor} \\ 
&=& 
  \!\!\!a^6 \sum_{\vec{x}, \vec{z}} \, \big\langle  T \left\{ J_{\nu}(\vec{0},
    t_f) J_{\mu}(\vec{z}, t_i)  P(\vec{x},t_0) \right\} \big\rangle \,
  e^{i \vec{p} \vec{x}} \, e^{-i \vec{q}_1 \vec{z}}.  \nonumber 
\eea
Here $\tau=t_i-t_f$ is the time separation between the two vector
currents and $t_{\pi}={\rm min}(t_f-t_0,t_i-t_0)$. The matrix element
in Eq.~(\ref{definition_TFF}) with an on-shell pion is obtained by
considering the limit of large $t_{\pi}$. With the definitions
\bea 
A_{\mu\nu}(\tau) & = & \lim_{t_{\pi} \rightarrow +\infty}
C^{(3)}_{\mu\nu}(\tau,t_{\pi}) \, e^{E_{\pi}t_{\pi}}, 
\label{A_munu}
\\ 
\widetilde{A}_{\mu\nu}(\tau) & = & \left\{\begin{array}{l@{~~~}l}
    A_{\mu\nu}(\tau) & \tau >0 \\  A_{\mu\nu}(\tau) \,  e^{-E_{\pi}
      \tau} & \tau<0 \end{array} \right.\;, 
\label{A_munu_tilde}
\eea 
one obtains 
\bea 
 M_{\mu\nu}^{\rm E} & = & \frac{2 E_{\pi}}{ Z_{\pi} }
 \int_{-\infty}^{\infty} \, \mathrm{d}\tau \, e^{\omega_1 \tau} \,
 \widetilde{A}_{\mu\nu}(\tau) \,.  
\label{lat_M} 
\eea 
The overlap factor $Z_{\pi}$ and the pion energy $E_{\pi}$ can be
obtained from the asymptotic behavior of the pseudoscalar two-point
function.   

For the calculation of $\FF(q_1^2,q_2^2)$ on the lattice in
Ref.~\cite{TFF_Lattice} we used eight CLS (Coordinated Lattice
Simulations) lattice ensembles~\cite{CLS_ensembles} with $n_f = 2$
dynamical quarks with three different lattice spacings $a = (0.048,
0.065, 0.075)~\fm$ and pion masses in the range $194 - 437~\MeV$.

\begin{figure*}[h!]

	\begin{minipage}[c]{0.49\linewidth}
	\centering 
	\includegraphics*[width=0.95\linewidth]{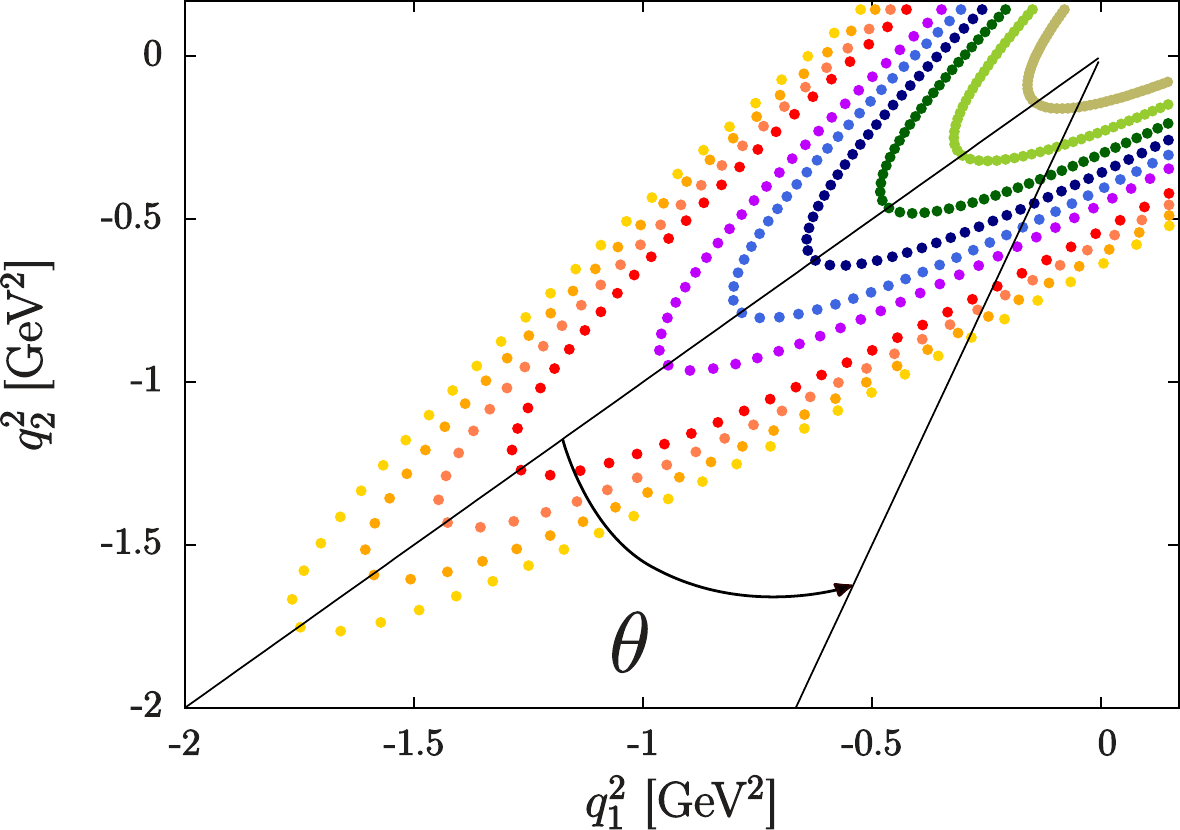}
	\end{minipage}
	\begin{minipage}[c]{0.49\linewidth}
	\centering 
        \vspace*{-0.2cm} 
	\includegraphics*[width=0.95\linewidth]{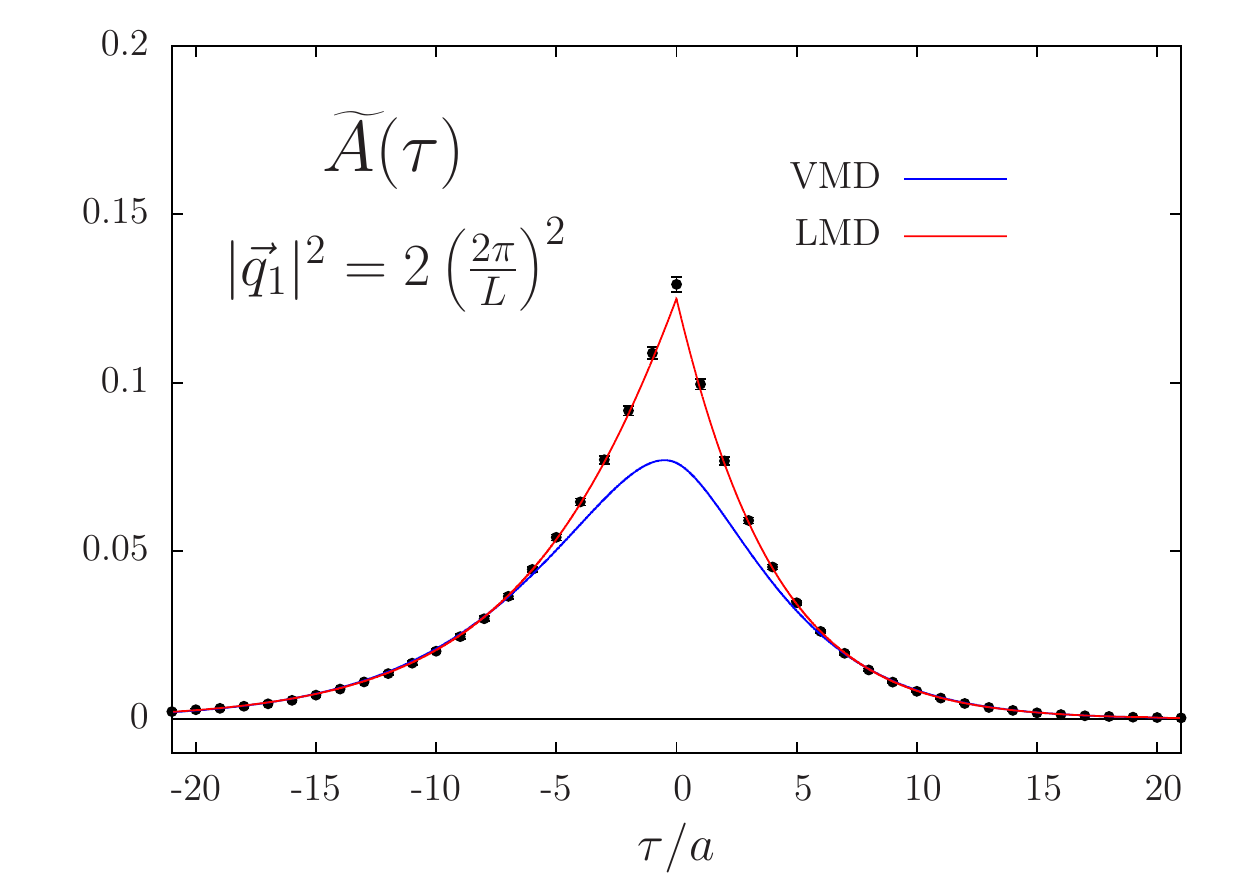}
	\end{minipage}

	\caption{(Left) Kinematic reach in the photon virtualities
          ($q_1^2,q_2^2$) with the pion at rest, for a lattice
          resolution of $64^3 \times 128$ with $a=0.048~\fm$ and
          $m_\pi = 268~\MeV$. (Right) The function
          $\widetilde{A}(\tau)$ (black points) and the VMD (blue line)
          and LMD (red line) fits used to describe the tail of the
          function at large $\tau$ for another lattice ensemble with
          $a=0.065~\fm$ and $m_\pi = 270~\MeV$.}	   
	\label{fig:kin}
\end{figure*}

We choose the pion rest frame $\vec{p} = 0$, with photons back-to-back
spatially $\vec{q}_2 = - \vec{q}_1$, where the kinematical range
accessible on the lattice is given by $q_1^2 = \omega_1^2 -
\vec{q}_1^{\, 2}$, $q_2^2 = (m_{\pi} - \omega_1)^2 - \vec{q}_1^{\,2}$.
With multiple values of $|\vec{q}_1|^2 = \left(\frac{2\pi}{L}\right)^2
|\vec{n}|^2$, $|\vec{n}|^2 = 1,2,3, \ldots$ we obtain mostly spacelike
photon virtualities up to $|q_{1,2}^2| \approx 1.5~\GeV^2$, as can be
seen in Fig.~\ref{fig:kin} (left). In practice, discrete values of
$\omega_1$ have been used to sample the momenta. Note that on the
lattice it is actually easier to access the double-virtual TFF
$\FF(q_1^2,q_2^2)$, in particular near the diagonal $q_1^2 = q_2^2$,
than the single-virtual form factor $\FF(q_{1,2}^2,0)$ along the two
axis, in contrast to the situation in experiments~\cite{TFF_exps}.

From the theoretical side, the form factor is constrained by the
chiral anomaly such that $\FF(0, 0) = 1/(4\pi^2
F_{\pi})$~\cite{adler_bell_anomaly} (in the chiral limit). For the
single-virtual form factor one expects the Brodsky-Lepage (BL)
behavior $\FF(-Q^2, 0) \xrightarrow[Q^2 \to \infty]{} 2 F_{\pi} /
Q^2$~\cite{BL_3_papers}. The precise value of the prefactor is,
however, under debate. The double-virtual form factor, where both
momenta become simultaneously large, has been computed using the OPE
at short distances. In the chiral limit the result reads $\FF(-Q^2,
-Q^2) \xrightarrow[Q^2 \to \infty]{} 2 F_{\pi} / (3 Q^2)$~\cite{OPE}.

In order to get a result for the double-virtual TFF $\FF(q_1^2,
q_2^2)$ in the continuum and for the physical pion mass, we fit our
lattice data, obtained for the eight ensembles with different lattice
spacings $a$ and pion masses $m_\pi$, with three simple models:
vector-meson dominance (VMD), lowest-meson dominance (LMD) and LMD+V,
see Refs.~\cite{LMD, LMD+V} for details about these models.  The often
used VMD model fulfills the BL behavior for the single-virtual case
(and describes quite well the available experimental data below
$2-3~\GeV^2$), but falls off too fast for the double virtual case
$\FF^{\rm VMD}(-Q^2, -Q^2) \sim 1/Q^4$. The LMD model is constructed
in such a way that it fulfills the OPE constraint, but fails to
reproduce the Brodsky-Lepage behavior. Finally, the LMD+V models is a
generalization of the LMD model and contains two vector resonances
$\rho$ and $\rho^\prime$. It can be made to fulfill both the BL and
the OPE constraints, for the prize of a large number of free
parameters.

In order to reduce the number of fit parameters for all the models, a
global fit is performed where all lattice ensembles are fitted
simultaneously assuming a linear dependence of each model parameter on
the lattice spacing $a$ and on the squared pion mass $m_\pi^2$, see
Ref.~\cite{TFF_Lattice} for all the details and results of the
fits.

The fits for the VMD and the LMD model are also used to perform the
integration in Eq.~(\ref{lat_M}) up to infinite $\tau$, see
Fig.~\ref{fig:kin} (right). For $|\tau| < 1.3~\fm$ the lattice data are
used. The dependence on these models for large $\tau$ is small, but
the behavior for small $\tau$ is very different. In fact, both the
lattice data and the LMD model show a cusp at $\tau = 0$, which is
related to the OPE in the double-virtual case, whereas the VMD model
is smooth at $\tau = 0$, since the VMD model falls off too fast at
large $Q^2$.

\begin{figure*}[h!]
	
	\begin{minipage}[c]{0.32\linewidth}
	\centering 
	\includegraphics*[width=0.95\linewidth]{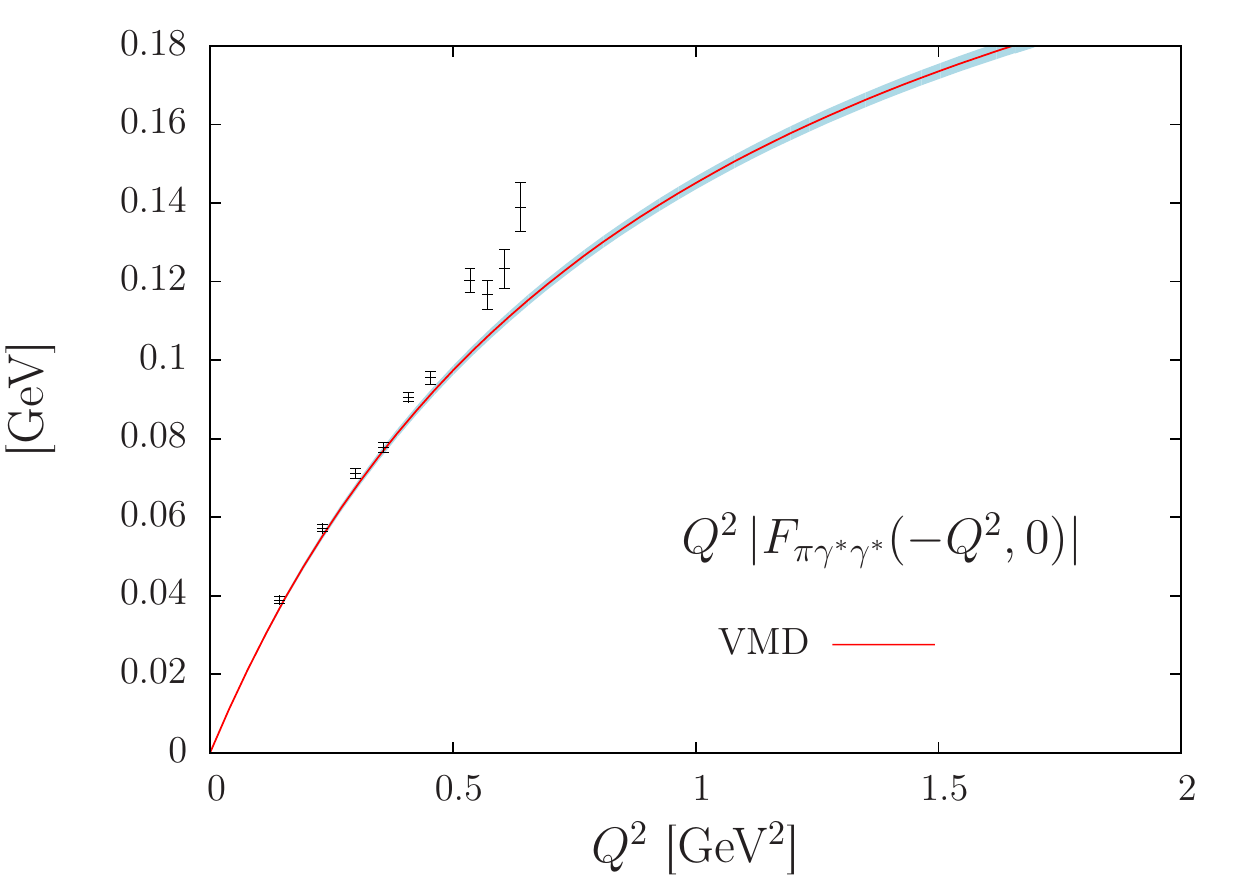}
	\end{minipage}
	\begin{minipage}[c]{0.32\linewidth}
	\centering 
	\includegraphics*[width=0.95\linewidth]{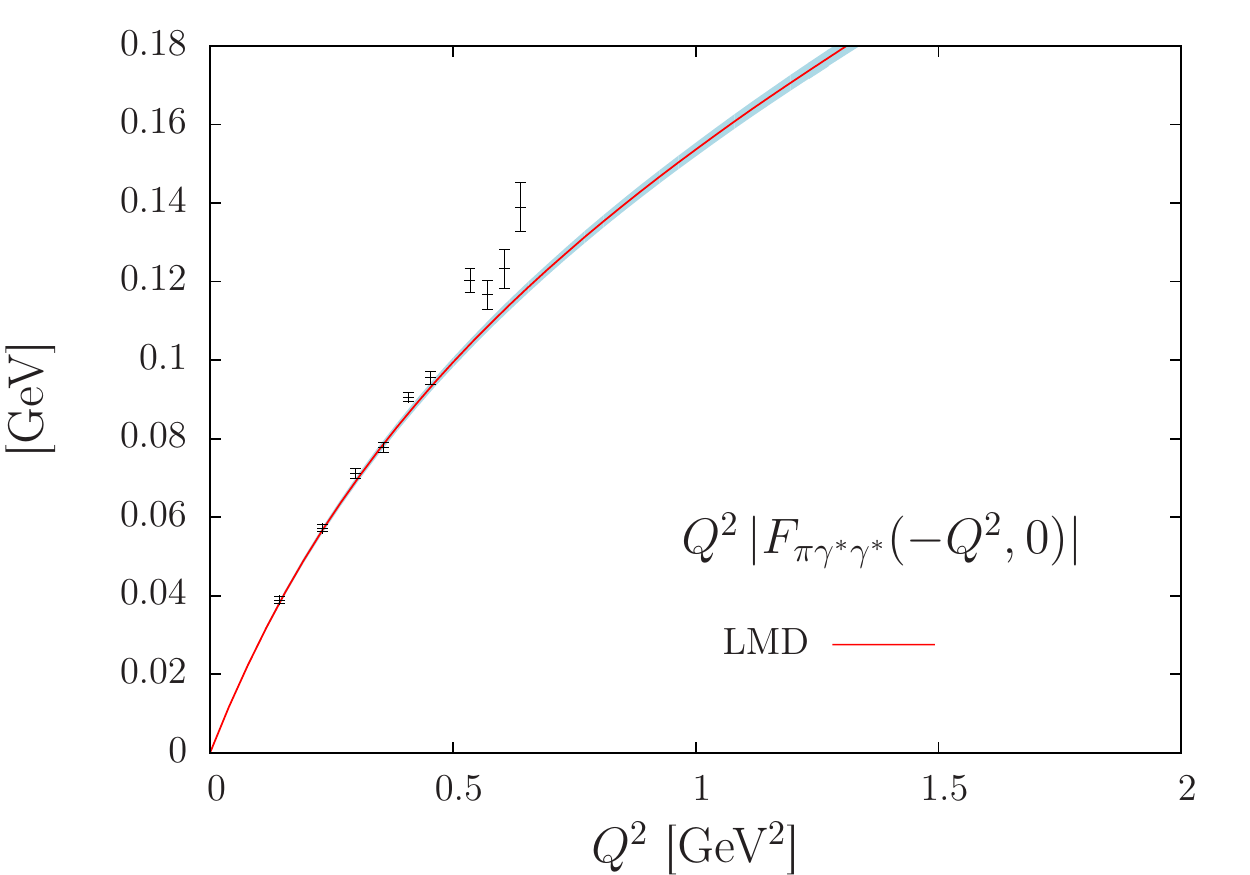}
	\end{minipage}
	\begin{minipage}[c]{0.32\linewidth}
	\centering 
	\includegraphics*[width=0.95\linewidth]{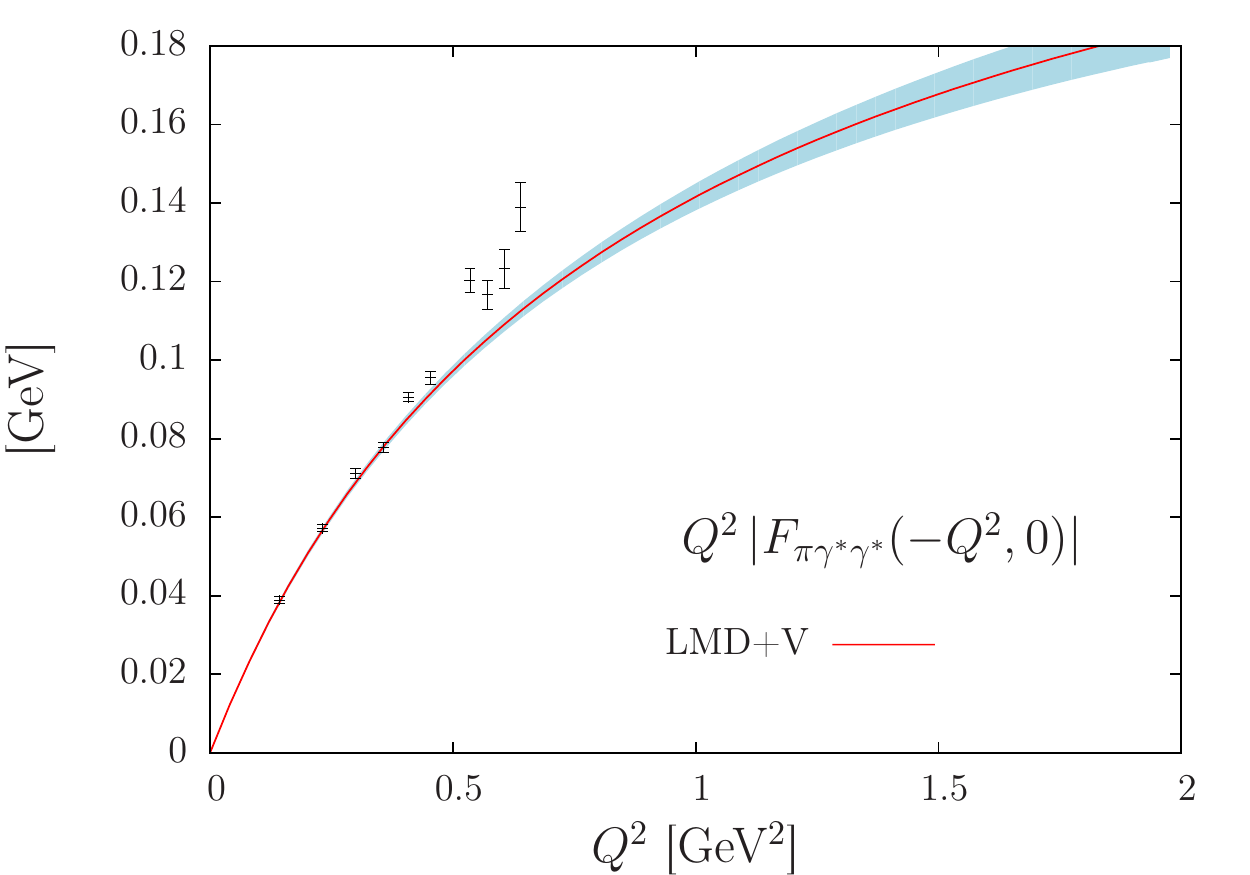}
	\end{minipage}

	\begin{minipage}[c]{0.32\linewidth}
	\centering 
	\includegraphics*[width=0.95\linewidth]{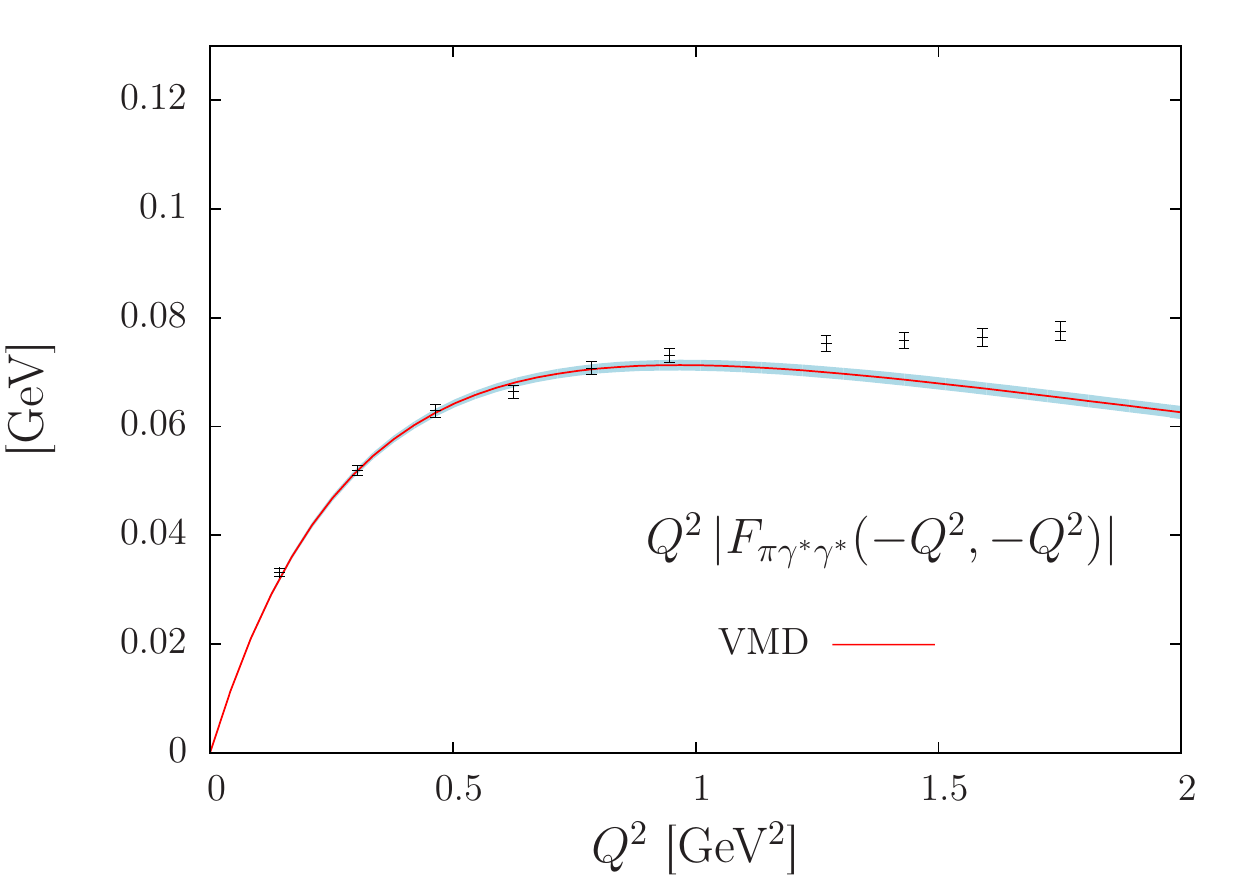}
	\end{minipage}
	\begin{minipage}[c]{0.32\linewidth}
	\centering 
	\includegraphics*[width=0.95\linewidth]{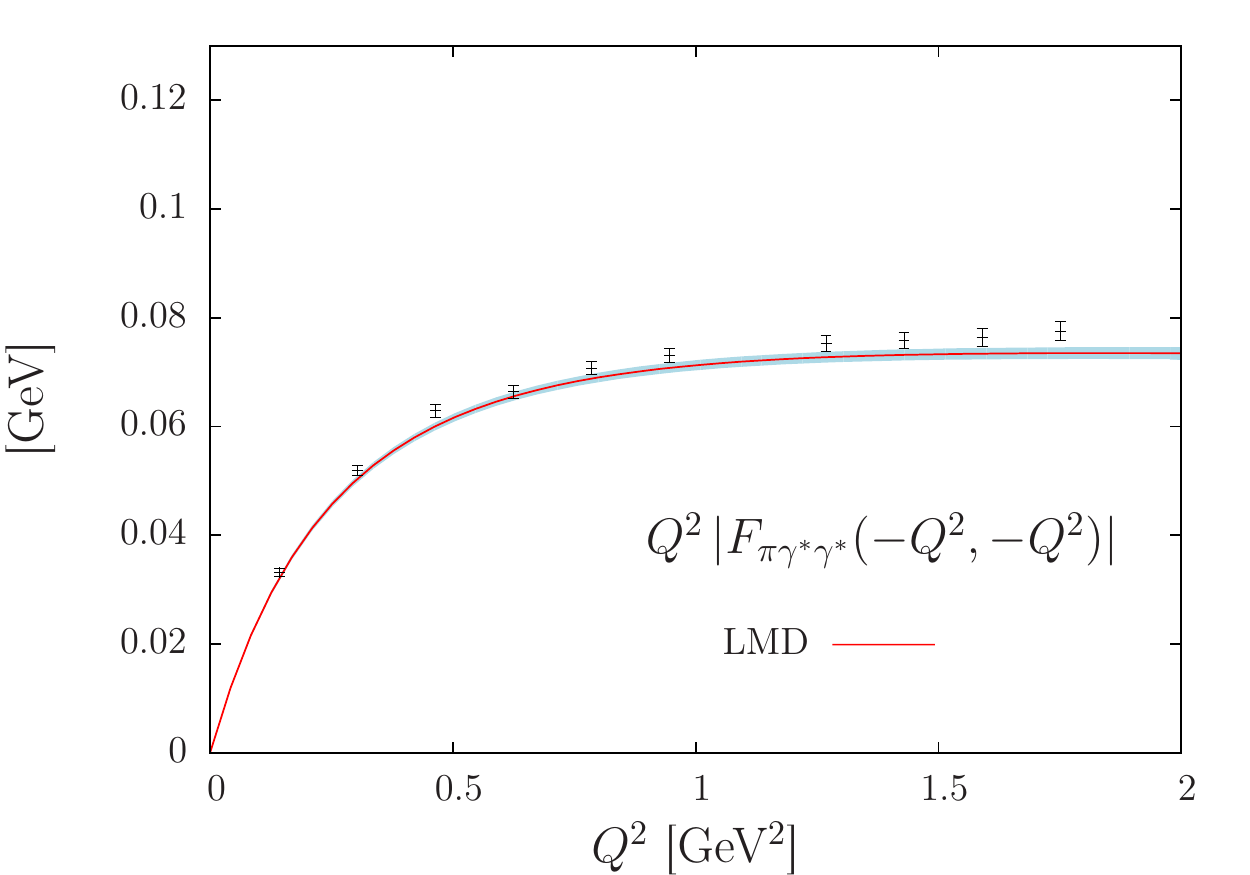}
	\end{minipage}
	\begin{minipage}[c]{0.32\linewidth}
	\centering 
	\includegraphics*[width=0.95\linewidth]{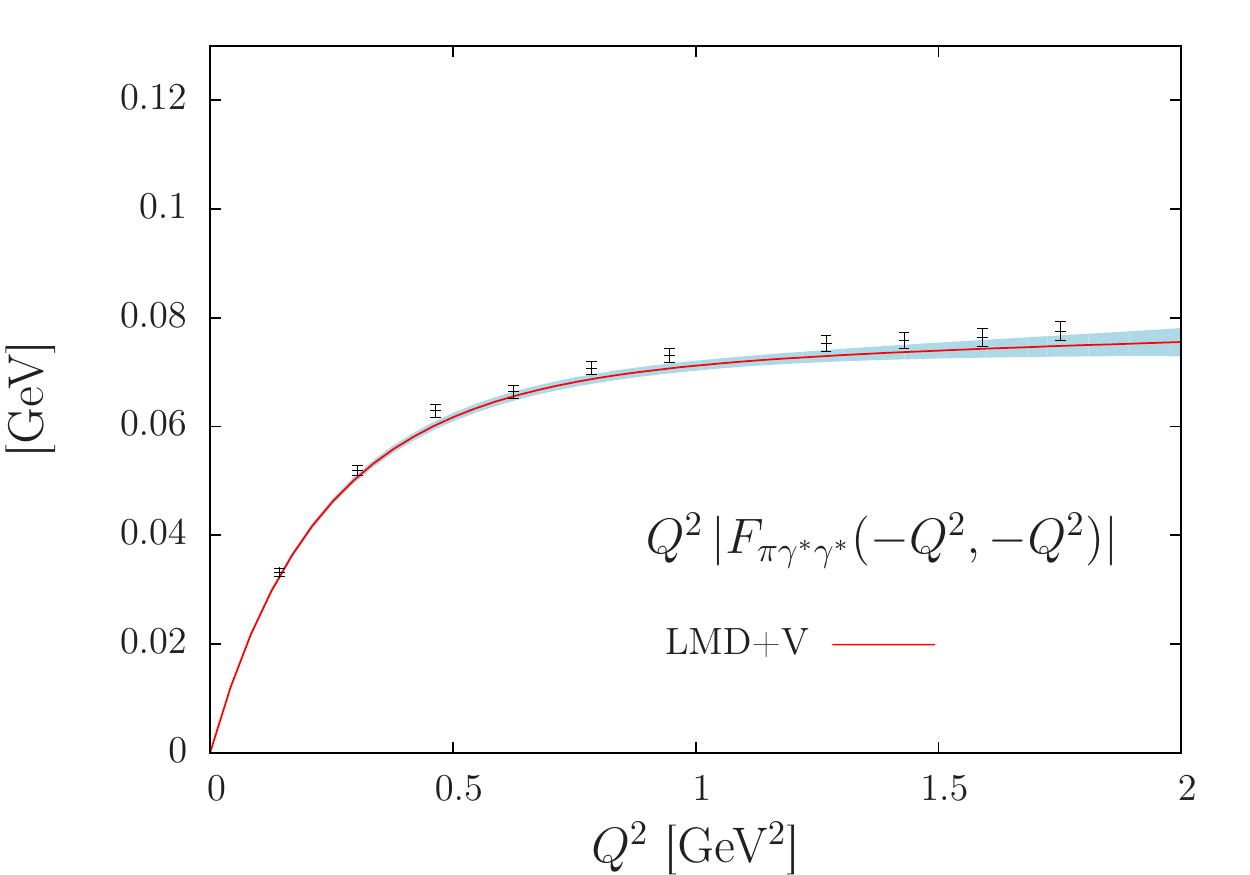}
	\end{minipage}
	
	\caption{Comparison of the VMD, LMD and LMD+V fits for a 
          lattice ensemble with $a = 0.048~\fm$ and $m_\pi =
          268~\MeV$. The red line corresponds to the results 
          from our global fit. Note that the points at different $Q^2$
          are correlated.}	   
	\label{fig:fit_ff}
\end{figure*}

The VMD model leads to a poor description of our data, with
$\chi^2/\dof = 2.9$ (uncorrelated fit), especially in the
double-virtual case and at large Euclidean momenta, see
Fig.~\ref{fig:fit_ff}. The normalization $\FF^{\rm VMD}(0,0)$ is off
from the value expected from the chiral anomaly and the fitted vector
meson mass does not agree with the $\rho$-mass. 

On the other hand, both the LMD model and the LMD+V model lead to a
quite good fit. The LMD model has a $\chi^2/\dof = 1.3$ (uncorrelated
fit) and leads to a determination of the chiral anomaly at the 7\%
level, with a value in agreement with expectations. The fitted vector
meson mass is close to the rho-meson mass. Furthermore, the fit result
for another parameter that is related to the OPE is compatible with
the theoretical expections. Despite the fact that the LMD model fails
to reproduce the BL behavior for the single-virtual form factor, this
does not seem to affect the global fit, since there are only few
lattice data points at rather low momenta in the single virtual case,
see Figs.~\ref{fig:kin} and \ref{fig:fit_ff}, i.e.\ one is not yet
sensitive to the asymptotic behavior. 

Finally, the LMD+V model has a $\chi^2/\dof = 1.4$ (uncorrelated fit),
however, only after the model parameters that are related to the two
vector meson masses, the BL behavior and the OPE constraint, have been
fixed, otherwise no stable fit was obtained. The chiral anomaly is
reproduced with 9\% accuracy and two other fitted model parameters are
close to results obtained in phenomenological analyses of the
TFF~\cite{LMD+V}.

\begin{figure*}[h!]

	\begin{minipage}[c]{0.49\linewidth}
	\centering 
	\includegraphics*[width=0.95\linewidth]{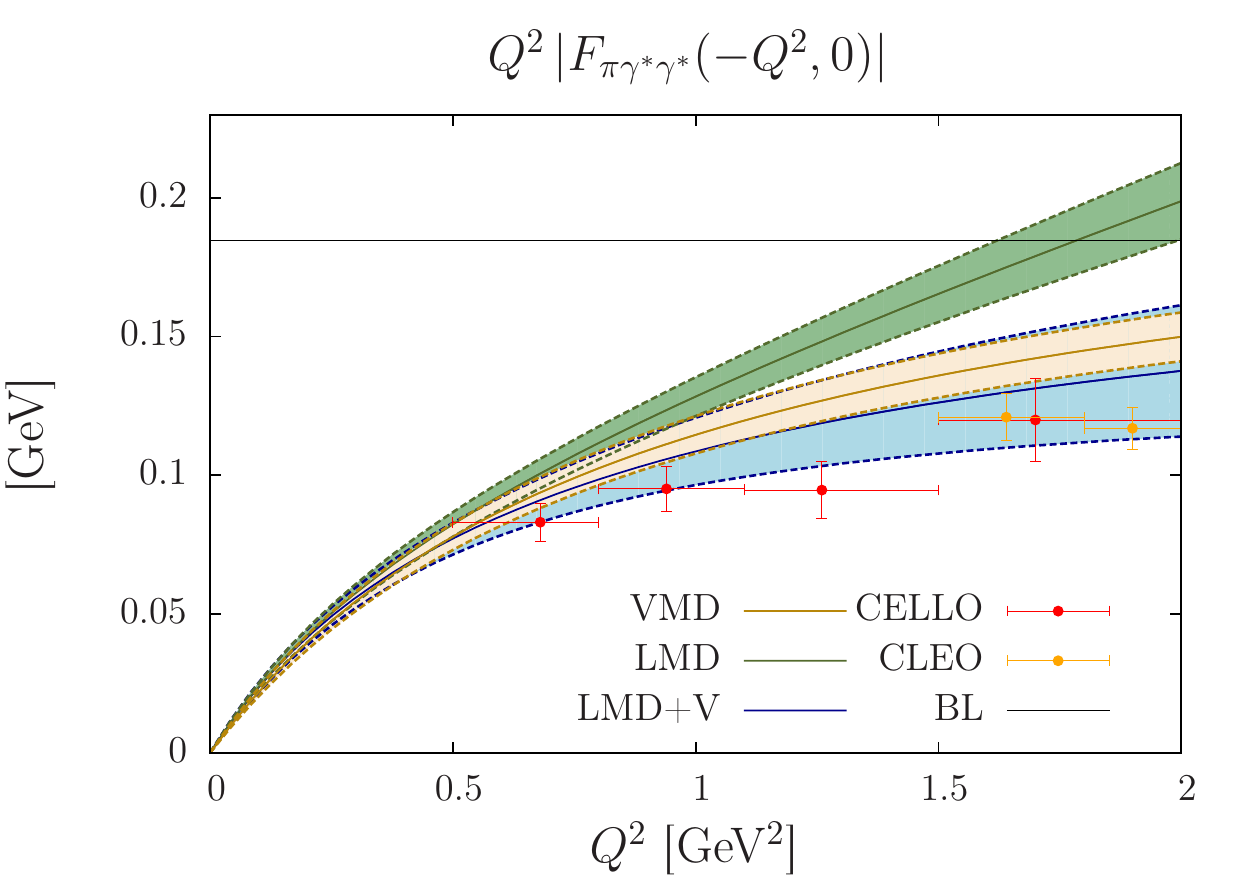}
	\end{minipage}
	\begin{minipage}[c]{0.49\linewidth}
	\centering 
	\includegraphics*[width=0.95\linewidth]{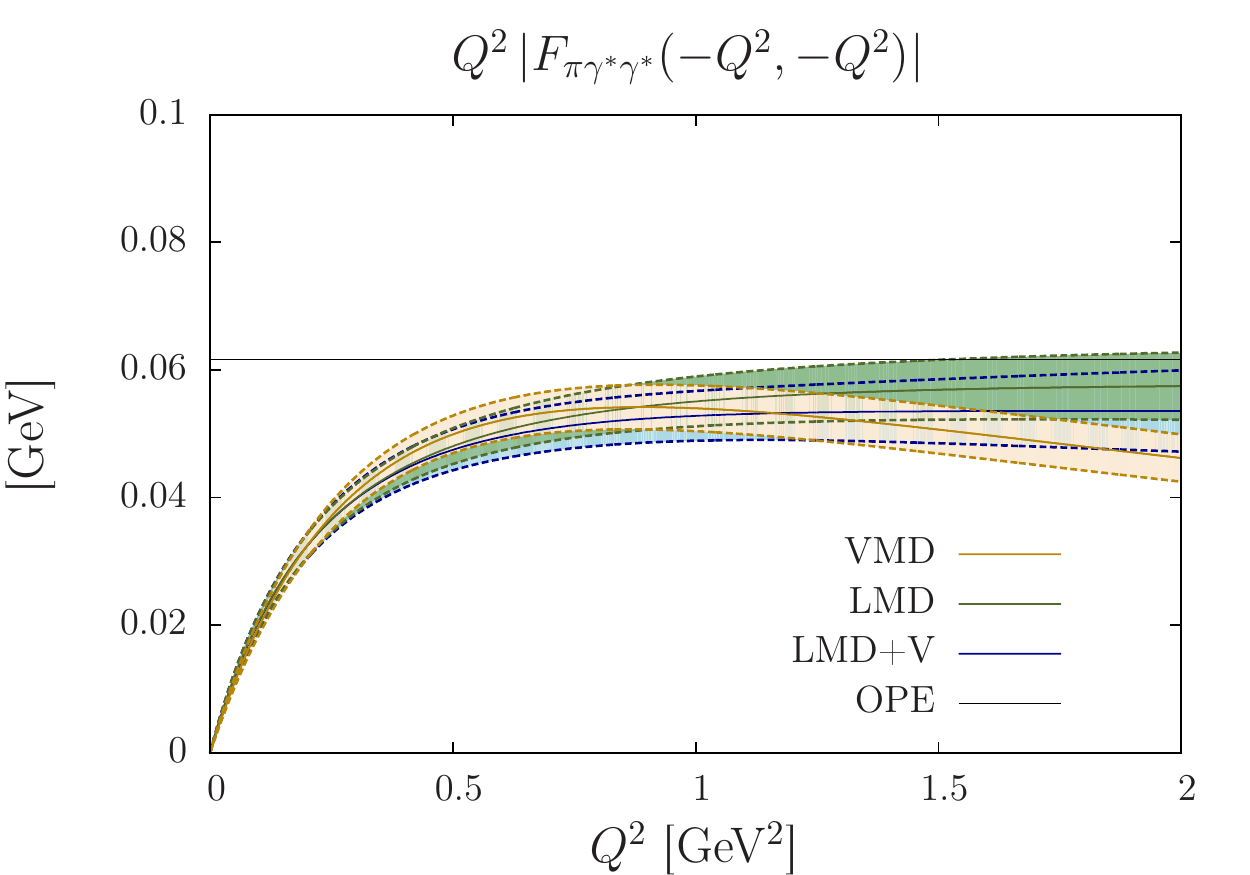}
	\end{minipage}
	
	\caption{Extrapolation of the lattice data to the continuum
          and the physical pion mass for the VMD, LMD and LMD+V
          models. (Left) Single-virtual form factor, compared with
          experimental results~\cite{TFF_exps} and the expectation for
          the asymptotic behavior according to
          Brodsky-Lepage~\cite{BL_3_papers}. (Right) Double-virtual
          form factor at $Q_1^2=Q_2^2$ and the expectation from the
          OPE~\cite{OPE}.} 
 	\label{fig:FF_cmp}
\end{figure*}

The form factors in the three models, extrapolated to the physical
point, are shown in Fig.~\ref{fig:FF_cmp}. In the single-virtual case,
the LMD+V model is in quite good agreement with the experimental data
from Ref.~\cite{TFF_exps}. The LMD model starts to deviate already at
$Q^2=1~\GeV^2$. In the double-virtual case, the LMD and LMD+V models
are quite similar and already close to their asymptotic behavior at
the largest point $Q^2 \sim 1.5~\GeV^2$ where we have lattice data.

Using the LMD+V model at the physical point from our fit, the
pion-pole contribution to HLbL in the muon $g-2$ can be obtained from
the three-dimensional integral representation from Ref.~\cite{JN_09},
with the result~\cite{TFF_Lattice},
\be \label{amupi0LMDV_lattice}
a_{\mu; \LMDV}^{\mathrm{HLbL}; \pi^0} = (65.0 \pm 8.3) \times 10^{-11} \,.  
\ee
Note that the given error is only statistical. No attempt has been
made to estimate the systematical errors from using different fit
models. Since the relevant momentum range in the pion-pole
contribution is below $1~\GeV$~\cite{KN_02_N_16}, i.e.\ where most of
our lattice data points are obtained, the extrapolation to large
momenta with the LMD+V or the LMD model does not have a large effect
on the final result.  The result in Eq.~(\ref{amupi0LMDV_lattice}) is
fully consistent with most model calculations which yield results in
the range $\amupi = (50-80) \times 10^{-11}$, but with rather
arbitrary, model-dependent error estimates, see Refs.~\cite{JN_09,
  Bijnens_16_17, KN_02_N_16} and references therein.

\clearpage 

\clearpage 

\section{HLbL forward scattering amplitudes in lattice QCD}

Using parity and time-reversal invariance of QCD, there are eight
independent light-by-light forward scattering amplitudes describing
the process $\gamma^*(\lambda_1,q_1) \, \gamma^*(\lambda_2,q_2) \to
\gamma^*(\lambda_1^\prime,q_1) \, \gamma^*(\lambda_2^\prime,q_2)$,
where $q_i$ and $\lambda_i^{(\prime)}=0,\pm$ are the momenta and
helicities of the virtual photons $(i=1,2)$. Six amplitudes are even
and two are odd functions of the crossing-symmetric variable $\nu =
q_1 \cdot q_2$. The forward scattering amplitudes
$\mathcal{M}_{\lambda_1^\prime\lambda_2^\prime\lambda_1\lambda_2}$ can
be related via the optical theorem to two-photon fusion amplitudes
$\mathcal{M}_{\lambda_1\lambda_2}$ for the process
$\gamma^*(\lambda_1,q_1) \, \gamma^*(\lambda_2,q_2) \to X(p_X)$ as
follows:
\bea 
\lefteqn{W_{\lambda_1^\prime \lambda_2^\prime, \lambda_1 \lambda_2} =     
\mathrm{Im}\,\mathcal{M}_{\lambda_1^\prime \lambda_2^\prime, \lambda_1
  \lambda_2}} \nonumber \\
& = & 
\frac{1}{2} \int \mathrm{d} \Gamma_X (2\pi)^4 \delta(q_1+q_2-p_X)
\nonumber \\ 
& & \quad \times \, 
\mathcal{M}_{\lambda_1 \lambda_2}(q_1,q_2,p_X) \,
\mathcal{M}^{*}_{\lambda_1^\prime \lambda_2^\prime}(q_1,q_2,p_X)
\,. \label{opt} 
\eea

Unitarity and analyticity then allow one to write dispersion relations
in $\nu$ at fixed values of the virtualities $Q_i^2 = -
q_i^2$. Performing one subtraction to get a faster convergence and
to suppress higher resonance states, one obtains the following sum rules
(omitting all dependence on the virtualities and the helicities):
\be \label{sr_even}
\mathcal{M}_{\rm even}(\nu) = \mathcal{M}_{\rm even}(0) + \frac{2
  \nu^2}{\pi} \int_{\nu_0}^\infty \! d \nu^\prime
\frac{W_{\rm even}(\nu^\prime)}{\nu^{\prime}(\nu^{\prime \, 2} - \nu^2-i
\epsilon) }  \,, 
\ee
\be \label{sr_odd}
\mathcal{M}_{\rm odd}(\nu) = \nu \mathcal{M}^{\prime}_{\rm odd}(0)
+ \frac{2 \nu^3}{\pi} \int_{\nu_0}^\infty \! d \nu^\prime \frac{W_{\rm
  odd}(\nu^\prime)}{\nu^{\prime 2} ( \nu^{\prime \, 2} - \nu^2-i
\epsilon) }  \,,   
\ee
with $\nu_0 = (Q_1^2 + Q_2^2)/2$. For later use, we introduce the
following notation for the subtracted even
$\mathcal{\overline{M}}(q^2_1,q^2_2,\nu) =
\mathcal{M}(q^2_1,q^2_2,\nu) - \mathcal{M}(q^2_1,q^2_2,0)$ and odd
$\mathcal{\overline{M}}(q^2_1,q^2_2,\nu) =
\mathcal{M}(q^2_1,q^2_2,\nu) - \nu
\mathcal{M}^{\prime}(q^2_1,q^2_2,0)$ amplitudes.

As proposed first in Ref.~\cite{HLbL_SR_Lattice_15} for the forward
scattering amplitude ${\mathcal{M}}_{TT}$ and extended recently to all
eight amplitudes in Ref.~\cite{HLbL_SR_Lattice_17}, the scattering
amplitudes on the left-hand sides of
Eqs.~(\ref{sr_even})-(\ref{sr_odd}) can be computed from the
correlation function of four vector currents on the lattice.  On the
other hand, the right-hand sides of
Eqs.~(\ref{sr_even})-(\ref{sr_odd}) are related to the two-photon
fusion processes using Eq.~(\ref{opt}). The latter are described by
single-meson TFFs which can be parametrized by simple models, as
sketched below. Fitting the lattice data then determines the model
parameters and the TFFs. This will allow one to calculate the
contributions from single-meson poles to the HLbL contribution in the
muon $g-2$.

\begin{figure*}
	
	\center
	\includegraphics*[width=0.49\linewidth]{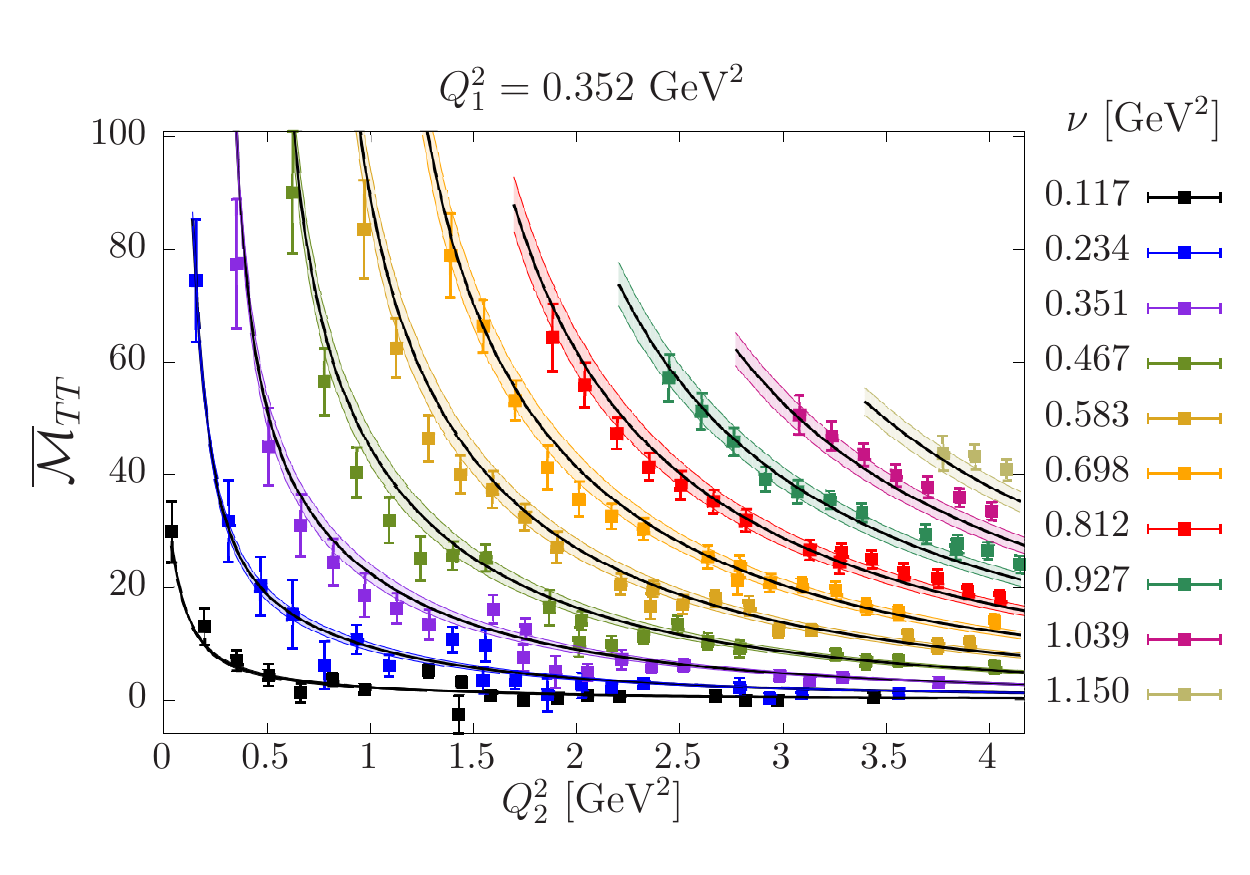}
	\includegraphics*[width=0.49\linewidth]{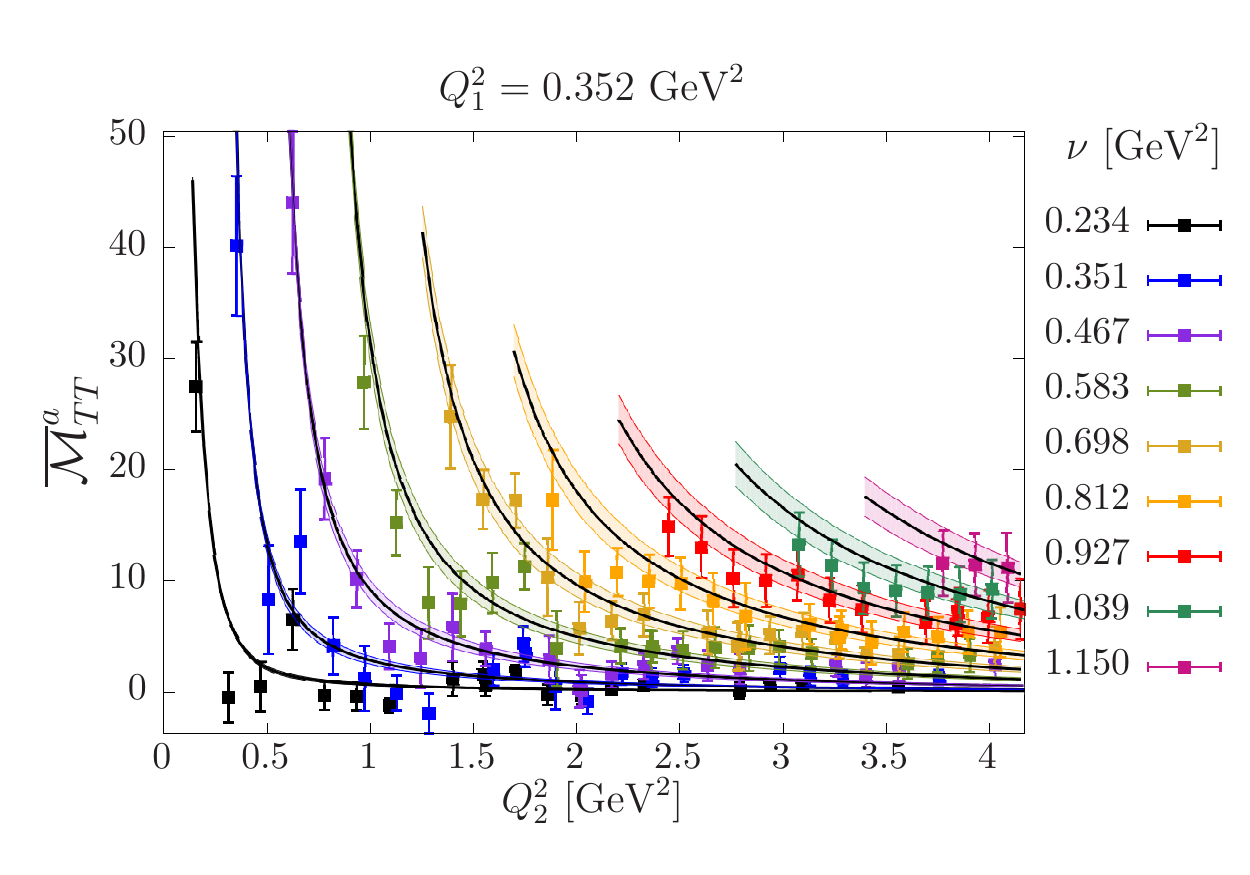}
        \\ 
        \vspace{-0.5cm} 
	\includegraphics*[width=0.49\linewidth]{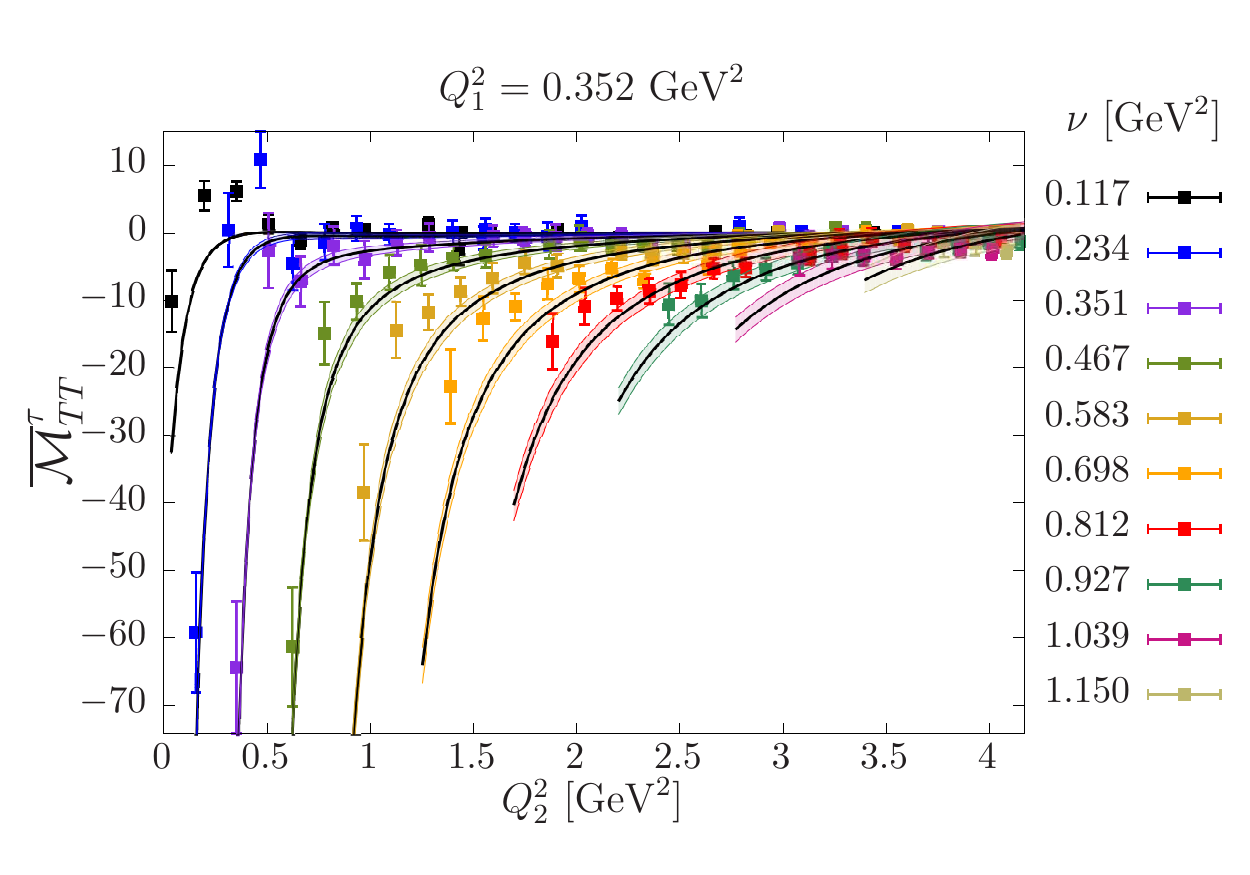}
	\includegraphics*[width=0.49\linewidth]{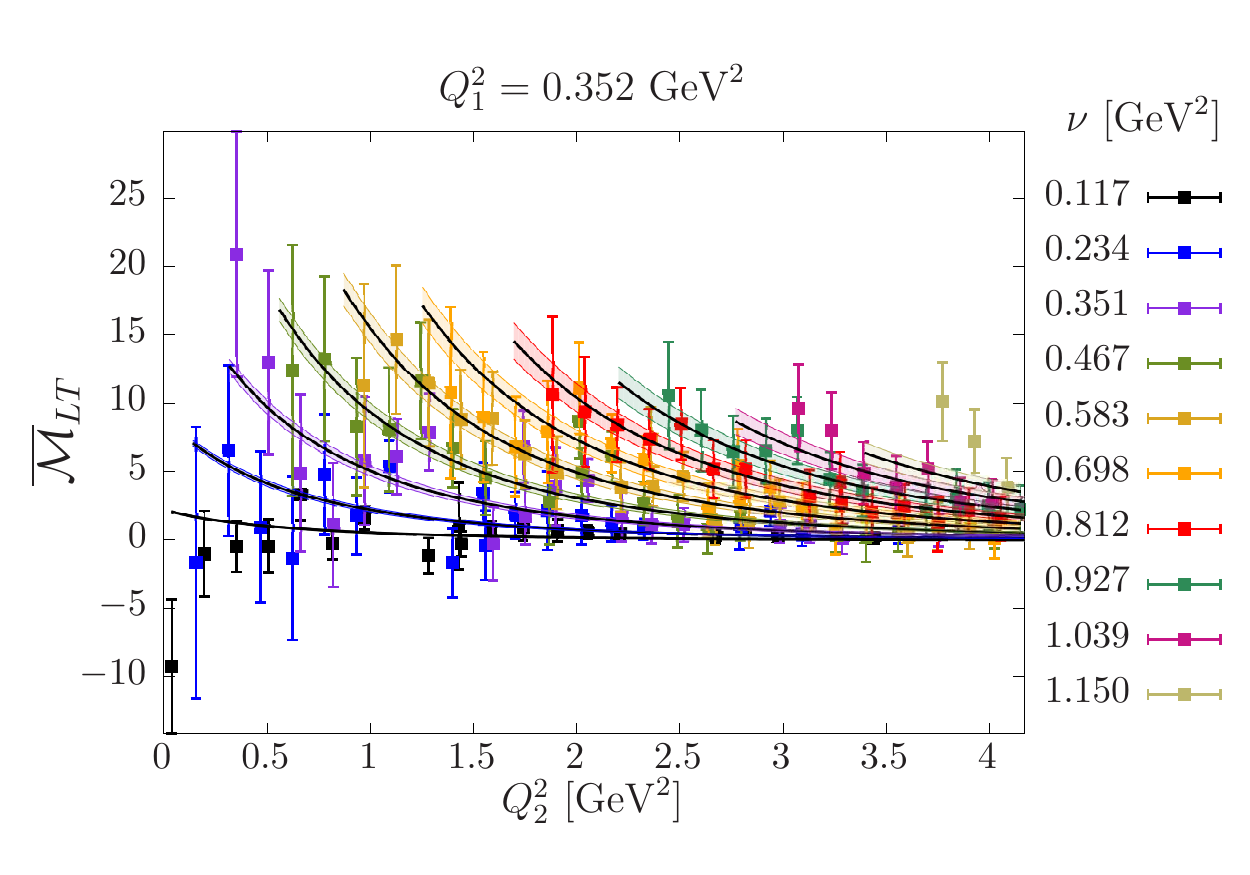}
        \vspace{-0.5cm} 

	\caption{The forward scattering amplitudes $\MsTT$, $\MsTTa$,
          $\MsTTt$ and $\MsLT$ $(\times10^6)$ for a lattice ensemble
          with $m_\pi = 314~\MeV$ and lattice spacing $a = 0.065~\fm$
          as function of $Q_2^2$ for a fixed $Q_1^2=0.352~\GeV^2$ and
          different values of $\nu$. The curves with error-bands
          represent the result of the fit of the phenomenological
          model described in the text to all eight scattering
          amplitudes with $\chi^2/\dof = 1.15$.    
          \label{fig:amp}}
\end{figure*}

The lattice simulations are performed on five CLS lattice
ensembles~\cite{CLS_ensembles} with two degenerate light dynamical
quarks at two lattice spacings $a = 0.048~\fm$ (one ensemble) and
$a=0.065~\fm$ (four ensembles) and pion masses in the range from
$194-437~\MeV$. For all ensembles the fully connected and for two
ensembles also the leading quark-disconnected diagrams contributing to
the four-point function are taken into account. For each ensemble, the
correlation function is computed at up to three values of $Q_1^2$
below $0.8~\GeV^2$ and for all values of $Q_2^2 \leq 4~\GeV^2$. The
result for a subset of four scattering amplitudes is shown in
Fig.~\ref{fig:amp}.

The photon fusion reaction on the right-hand sides of
Eqs.~(\ref{sr_even})-(\ref{sr_odd}) can produce any C-parity-even
state $X$. The main contribution is expected from the pseudoscalar
($0^{-+}$), scalar ($0^{++}$), axial-vector ($1^{++}$) and tensor
($2^{++})$ mesons, where we consider in each channel only the lightest
state. We are working with two degenerate dynamical quarks and fit our
phenomenological model to only the fully-connected diagrams. To
compensate for this, we include only the contributions from isovector
mesons, multiplied by a factor of
$34/9$~\cite{Bijnens_16_17,HLbL_SR_Lattice_17}.

There is one TFF for the pseudoscalars, two each for the scalars and
axial-vectors and four for the tensor
mesons~\cite{HLbL_sum_rules}. For the pseudoscalars, we take the TFF
from Ref.~\cite{TFF_Lattice} evaluated on the same lattice
ensembles. The other TFFs are parametrized as follows
\be \label{TFF_ansatz}
F_X(Q_1^2,Q_2^2) =  \frac{ F_X(0,0) }{ \left( 1 + Q_1^2/\Lambda_X^2
  \right)^n \left( 1 + Q_2^2/\Lambda_X^2 \right)^n } \,, 
\ee
where we assume a monopole ansatz $(n=1)$ for the scalars and a dipole
ansatz $(n=2)$ for the axial-vectors and tensor mesons, parametrized
by the mass scale $\Lambda_X$. We assume one common mass for the
scalar TFFs, one mass for the TFFs of the axial-vectors and four
different masses for the TFFs of the tensor mesons. These six masses
will be considered as free fit parameters.

The normalization of the TFFs is given by the two-photon decay width
and is taken from experiment where available, e.g.\ for the scalars
$\GammaGG = \frac{\pi \alpha^2}{4} m_S \left[ F^T_{{\cal S}
    \gamma^\ast \gamma^\ast}(0, 0) \right]^2 $ (an appropriately
defined effective two-photon width is employed for the
axial-vectors~\cite{HLbL_sum_rules}). Since not all normalizations
have been measured, further input from dispersive sum rules from
Ref.~\cite{Danilkin_Vanderhaeghen_17} is used. Furthermore, for the
pseudoscalars, again the lattice data from Ref.~\cite{TFF_Lattice} are 
used.

As an example, the pseudoscalar contribution to the cross-section
$\sigma_{TT}$ of two transversely polarized photons is given in the
narrow-width approximation by 
\be
\sigma_{TT} = 8 \pi^2 \delta(s-m_P^2) \frac{\GammaGG}{m_P}  \frac{
  2\sqrt{X} }{ m_P^2 } 
\left[ \frac{ F_{{\cal P} \gamma^\ast \gamma^\ast}(Q_1^2, Q_2^2)
     }{ F_{{\cal P} \gamma^\ast \gamma^\ast}(0, 0)  } \right]^2 \,, 
\ee
where $X = \nu^2 - Q_1^2 Q_2^2$ is the virtual-photon flux
factor. Similar results can be obtained for the other mesons
where we assume a Breit-Wigner shape for the
resonances~\cite{HLbL_SR_Lattice_17}. 

In Fig.~\ref{fig:amp} the results for four scattering amplitudes of a
combined fit to all eight amplitudes with the phenomenological model
described above is shown for one lattice ensemble. For four ensembles,
the $\chi^2/\dof$ is quite good, between $1.13-1.35$. The fit for the
ensemble with the heaviest pion mass $m_\pi = 437~\MeV$ is, however,
not very satisfactory and that ensemble is left out in the chiral
extrapolation to the physical pion mass.

The relative contribution of the different mesons to the individual
scattering amplitudes was also studied in
Ref.~\cite{HLbL_SR_Lattice_17}. The pseudoscalar and tensor mesons
give the dominant contribution to the amplitudes $\MsTT$, $\MsTTt$ and
$\MsTTa$ that involve two transverse photons. The pseudoscalar meson
does not contribute to $\MsTL$, $\MsLT$ where the main contribution
are from axial and tensor mesons. In the amplitudes $\MsTLa$,
$\MsTLt$, $\MsLL$ scalar, axial and tensor mesons contribute
significantly. The contribution from $\gamma^* \gamma^* \to \pi^+
\pi^-$, evaluated with scalar QED dressed with a monopole vector form
factor, is always small compared to the other channels.

The lattice simulations are performed for ensembles away from the
physical quark masses. The pion and $\rho$-meson masses are set to
their lattice values obtained from the exponential decay of the
pseudoscalar and vector two-point functions. For other resonances, we
assume a constant shift in the masses $m_X = m_X^{\rm phys} +
(m_{\rho}^{\rm lat} - m_{\rho}^{\rm exp})$. In
Table~\ref{tab:ch_extrap} we compare the results of the masses in the
TFFs, extrapolated to the physical pion mass, to experimental and
phenomenological determinations~\cite{Resonance_masses_exps,
  Danilkin_Vanderhaeghen_17}. The agreement is reasonably good for
$M_S$, $M_T^{(2)}, M_T^{(0,L)}$, although the scalar mass on the
lattice is a bit high. There are quite strong tensions for $M_A,
M_T^{(1)}, M_T^{(0,T)}$. In particular the latter two tensor masses on
the lattice are almost a factor two larger than the phenomenological
determinations. See Ref.~\cite{HLbL_SR_Lattice_17} for a more detailed
discussion and potential reasons for the disagreements. Note in
particular, that we have not yet performed the continuum limit.

Overall, we get a good description of the lattice data with the
lattice determination of the pion TFF~\cite{TFF_Lattice} and the
simple monopole or dipole ans\"atze from Eq.~(\ref{TFF_ansatz}) for
the various TFF's with one resonance in each channel.

\begin{table}
\centering 
\caption{Chiral extrapolation to the physical pion mass for the scalar
  monopole mass $M_S$, the axial dipole mass $M_A$ and the four tensor
  dipole masses corresponding to different helicities, compared to
  experimental or phenomenological determinations. All masses are
  given in GeV. The ensembles have a finite lattice spacing of $a =
  0.065~\fm$ and the ensemble with the largest pion mass has been
  excluded in the chiral extrapolation.}     
\label{tab:ch_extrap}
\begin{tabular}{l@{~~~~~}c@{~~~~~}c}
\hline
             &  Lattice   & Experiment \\
\hline
$M_S$        &   1.04(14) & 0.796(54) \\
$M_A$        &   1.32(07) & 1.040(80) \\
$M_T^{(2)}$   &   1.35(24) & 1.222(66) \\
$M_T^{(1)}$   &   1.69(16) & 0.916(20) \\
$M_T^{(0,T)}$ &   1.96(09) & 1.051(36) \\
$M_T^{(0,L)}$ &   0.67(19) & 0.877(66) \\
\hline
\end{tabular}
\end{table}


\section{Conclusions}

Lattice QCD can provide a model-independent, first-principle
calculation of the HLbL contribution to the muon $g-2$. Recent first
preliminary results by RBC-UKQCD~\cite{HLbL_Lattice_RBC-UKQCD} and the
lattice group at Mainz~\cite{HLbL_Lattice_Mainz_15,
  HLbL_Lattice_Mainz_16, HLbL_Lattice_Mainz_17, TFF_Lattice,
  HLbL_SR_Lattice_15, HLbL_SR_Lattice_17} look very
promising. Hopefully, in a few years time when the final result from
the Fermilab experiment will be published, an estimate with 10\%
uncertainty (combined statistical and controlled systematics errors)
can be reached, which would match the expected experimental precision,
if one assumes that the central value for HLbL is close to current
model estimates $\amuhlbl \approx 100 \times 10^{-11}$~\cite{PdeRV_09,
  N_09, JN_09, Jegerlehner_15_17}.

Since HLbL involves a rank-four tensor with many independent momenta
(or space-time points in position space), it is a very complicated
object. We therefore think it makes sense to have as many tests and
observables as possible on the lattice, not just the final number for
$\amuhlbl$ itself. Therefore the calculation of the pion (light
pseudoscalar) transition form factor and of HLbL forward scattering
amplitudes, combined with HLbL sum rules, will be a valuable tool to
compare different lattice calculations, once they become available.


\begin{acknowledgement}
  We are grateful for the use of CLS lattice ensembles. We acknowledge
  the use of computing time on the JUGENE and JUQUEEN machines located
  at Forschungszentrum J\"ulich, Germany. The correlation functions
  were computed on the clusters ``Wilson'' at the Institute of Nuclear
  Physics, University of Mainz, ``Clover'' at the Helmholtz-Institute
  Mainz and ``Mogon'' at the University of Mainz, Germany.  This work
  was partially supported by the Deutsche Forschungsgemeinschaft (DFG)
  through the Collaborative Research Center ``The Low-Energy Frontier
  of the Standard Model'' (SFB 1044).
\end{acknowledgement}


\end{document}